\documentclass[twocolumn,pra]{revtex4}
\usepackage{graphicx}
\newcommand{\ket}[1]{|{#1}\rangle}			
\newcommand{\bra}[1]{\langle{#1}|}

\begin{document}
\title{Quantum circuit for security proof of quantum key distribution without encryption of error syndrome and noisy processing}

\author{Kiyoshi Tamaki$^{1,2}$}
\email{tamaki@will.brl.ntt.co.jp}
\author{Go Kato$^{3}$}
\affiliation{$^{1}$NTT Basic Research Laboratories, NTT Corporation,\\
3-1,Morinosato Wakamiya Atsugi-Shi, Kanagawa, 243-0198, Japan\\ 
$^{2}$CREST, JST Agency, 4-1-8 Honcho, Kawaguchi, Saitama, 332-0012, Japan\\
$^{3}$NTT Communication Science Laboratories, NTT Corporation\\
3-1,Morinosato Wakamiya Atsugi-Shi, Kanagawa, 243-0198, Japan
}

\begin{abstract}
One of the simplest security proofs of quantum key distribution is based on the so-called complementarity scenario, which involves the complementarity control of an actual protocol and a virtual protocol [M. Koashi, e-print arXiv:0704.3661 (2007)]. The existing virtual protocol has a limitation in classical postprocessing, i.e., the syndrome for the error-correction step has to be encrypted. In this paper, we remove this limitation by constructing a quantum circuit for the virtual protocol. Moreover, our circuit with a shield system gives an intuitive proof of why adding noise to the sifted key increases the bit error rate threshold in the general case in which one of the parties does not possess a qubit. Thus, our circuit bridges the simple proof and the use of wider classes of classical postprocessing.
\end{abstract}



\maketitle

\section{Introduction}
Quantum key distribution (QKD) is one of the most attractive research areas in quantum information theory, which has practical applications, and it has been intensively investigated both experimentally and theoretically. So far, many experiments of QKD have been reported (see, for instance, \cite{Experiments}), and some of them have demonstrated the actual distillation of the secret key \cite{SECOQC}. Since very hard work is needed for the actual implementation of QKD, it would be good from an experimental viewpoint if we could implement QKD with easier setups and easier classical postprocessing parts.

From the theoretical point of view, it is very important and interesting to consider a security proof in a simple manner. One of the simplest approaches is to consider a complementarity control of an actual protocol and a virtual protocol \cite{Koashi07}, which the sender (Alice) and the receiver (Bob) can choose to execute, but cannot execute simultaneously. In the actual protocol, the goal is to agree on the bit values along the computational basis, say the $Z$ basis while, in the virtual protocol, Alice and Bob collaborate to create an eigenstate of the $X$ basis (the conjugate basis of $Z$) in Alice's side. With these protocols, Koashi proved in \cite{Koashi07} that the necessary and sufficient condition for the secure key distillation is to successfully execute these complementary tasks. 

In this security proof, the virtual protocol has to be constructed in such a way that an adversary (Eve) cannot discriminate it from the actual one, and the existing virtual protocol assumes the encryption of the syndrome for the error-correction step \cite{Koashi05}. Since the classical postprocessing is very important in QKD, it is interesting to consider how to apply the complementarity control to QKD without the encryption, and even with wider classes of classical postprocessing. An interesting type of postprocessing is the so-called noisy processing, which was first introduced by Renner, {\it et. al} \cite{Renner05}. In this processing, by intentionally adding noise to one of the partiesfsifted key, the bit error rate threshold increases, and it is interesting to consider how to explain this processing from the viewpoint of the complementarity control.  


In this paper, we remove the encryption by explicitly constructing a qubit-based quantum circuit for the virtual protocol, and we employ our circuit with a shield system to give an intuitive proof of why the noisy processing increases the bit error rate threshold. Thus, our circuit bridges the simple security proof and the use of wider classes of classical postprocessing. One of the features of our quantum circuit is that it can output the syndrome, apply bit-flip operations, discard any unnecessary qubit, and output the secret key simultaneously. Thus, the virtual protocol with our quantum circuit can be equivalently converted to the actual protocol, and our circuit can accommodate the use of one-way and bi-directional error-reconciliation codes. 


Our approach to noisy processing assumes that only one of the parties has a qubit. This is one of the advantages over the original proposal \cite{Renner05} or private state distillation approach \cite{Renes07}, where both of the parties have to possess a qubit. We note that we can apply our quantum circuit to the security proof of protocols, such as BB84 \cite{BB84}, six-state protocol \cite{Six, note6}, and other protocols where the so-called phase error rate in the code qubits can be tightly estimated, i.e., Bob can guess Alice's fictitious $X$-basis bit value with arbitrary small failure probability as the size of the sifted key increases.


The organization of this paper is as follows. In Sec. \ref{review}, we briefly review the security proof based on the complementarity scenario \cite{Mayers96, Koashi07}, and then we construct the quantum circuit for a virtual protocol in Koashi's proof in Sec. \ref{our circuit}. Next, we apply our quantum circuit to the following cases: (i) Alice has a fictitious qubit in Sec. \ref{first case}, (ii) both Alice and Bob have fictitious qubits in Sec. \ref{second case}, and (iii) noisy processing in Sec. \ref{noisy processing}. Finally, we summarize our paper.

\section{Review of complementarity scenario}\label{review}
A way to prove the security of QKD is to consider a virtual protocol that is equivalent to the actual protocol and easy to analyze. Here, by ``equivalent'', we mean that the resulting secret key is the same between two protocols and that all the information available to Eve, including classical and quantum information, is the same. The former requirement is the equivalence from the users' view, and the latter is the one from Eve's view. Thus, the security proved in the virtual protocol means the security of the actual protocol. One approach along this line is to consider a virtual protocol that is based on the distillation of the Bell state \cite{EDP, LC98, SP00} of the form $\ket{\phi^{+}}\equiv\frac{1}{\sqrt{2}}(\ket{0_z}\ket{0_z}+\ket{1_z}\ket{1_z})$, which we call a Shor-Preskill type of proof. In this approach, the final key is generated via $Z$-basis (spanned by $\{\ket{0_z}, \ket{1_z}\}$) measurement, and the high fidelity of the distilled state relative to $\ket{\phi^{+}}$ guarantees the security since $\ket{\phi^{+}}$ is decoupled with Eve's system.

This proof can be seen from a different aspect. Since $\ket{\phi^{+}}$ has no probability of having the bit (bit in $Z$-basis) and phase (bit in $X$-basis: $\{\ket{0_x}\equiv(\ket{0_z}+\ket{1_z})\sqrt{2}, \ket{1_x}\equiv(\ket{0_z}-\ket{1_z})\sqrt{2}\}$) errors, one may conjecture that if Bob can perfectly predict Alice's measurement outcome regardless of which basis was chosen (of course, these two tasks cannot be performed simultaneously), Alice and Bob can share the secret key. This is an intuitive idea behind the complementarity control \cite{Mayers96}, and in \cite{Koashi07} Koashi formally introduced complementarity control of the actual protocol and the virtual protocol for the security proof. In the actual protocol, Alice and Bob try to share the same bit values in $Z$-basis, and Alice tries to generate a $X$-basis eigenstate with the help of Bob over the quantum channel in the virtual protocol. From the viewpoint of the Shor-Preskill type proof, the former (latter) protocol is related with the fact that Bob can guess Alice's bit value in $Z$-basis ($X$-basis) if Alice and Bob share $\ket{\phi^{+}}$. The actual protocol and virtual protocols are exclusive of each other, and in order to prove the security of the actual protocol, we require in the virtual protocol that Alice's and Bob's operations must commute with the measurement of the final keys, which is called nondisturbing condition. Then, Koashi proved that the task of the secret key distillation is equivalent to the complementarity control of the two protocols, and the security of the final key can be analyzed by the virtual protocol only.

As an explicit example of the virtual protocol, Koashi proposed the virtual protocol with the encryption of the syndrome \cite{Koashi05}. Since the encryption gives Eve no information, Alice and Bob can behave quite differently between the actual and virtual protocols. More precisely, Alice and Bob can choose any code, including nonlinear codes, for the error-correction in the actual protocol while they can perform any non-disturbing operations for Alice to prepare the $X$-basis eigenstate in the virtual protocol. This means that we do not need to consider the measurement of the syndrome in the virtual protocol, which makes the security analysis easy but limits classical postprocessing parts.

\section{Our circuit for the virtual protocol}\label{our circuit}

In order to remove the assumption of the encryption, we propose a quantum circuit that outputs both the syndrome for any linear codes and the required information for the distillation of the $X$-basis eigenstate. Since our circuit assumes a possession of qubits, it can be applied to any party who has a fictitious qubit state, such as Alice or a party with the squashing operator \cite{squash}, with the detector decoy \cite{MCL08}, with a photon number resolving detector, and with other techniques to define a qubit \cite{Koashi04}. For the explanation, we assume that Alice has the fictitious qubit, and through Alice's $Z$-basis measurement, the sifted key is determined. Moreover, we concentrate on the sifted key, and we consider only Alice's side in order to show that Alice's key is independent of Eve's system. The actual protocol runs as follow.

{\it Actual protocol:}\newline
(Step 1) Alice conducts measurements on her $(n+s)$ qubits, and she is left with $(n+s)$ bits of a sifted key.\newline
(Step 2) Alice computes $s$-bit syndrome $S^{(A)}_{z}$ for the error-reconciliation protocol to be sent to Bob over a public channel without encryption. Then, depending on the error-correction codes and the syndrome from Bob, Alice discards appropriate $s$ bits and applies bit-flip operations on the remaining $n$ bits. At this point, Alice has an $n$-bit reconciled key $\kappa_{\rm rec}^{(A)}$.\newline
(Step 3) Over a public discussion, Alice and Bob agree on randomly chosen independent $(n-m)$ $n$-bit strings $\{{\vec V}_{k}\}_{k=1,...,n-m}$. Alice takes $\{\kappa_{\rm rec}^{(A)}\cdot{\vec V}_{k}\}_{k=1,...,n-m}$ as the final key $\kappa_{\rm fin}^{(A)}$. 

\begin{figure}[tbp]
\begin{center}
\includegraphics[scale=0.25]{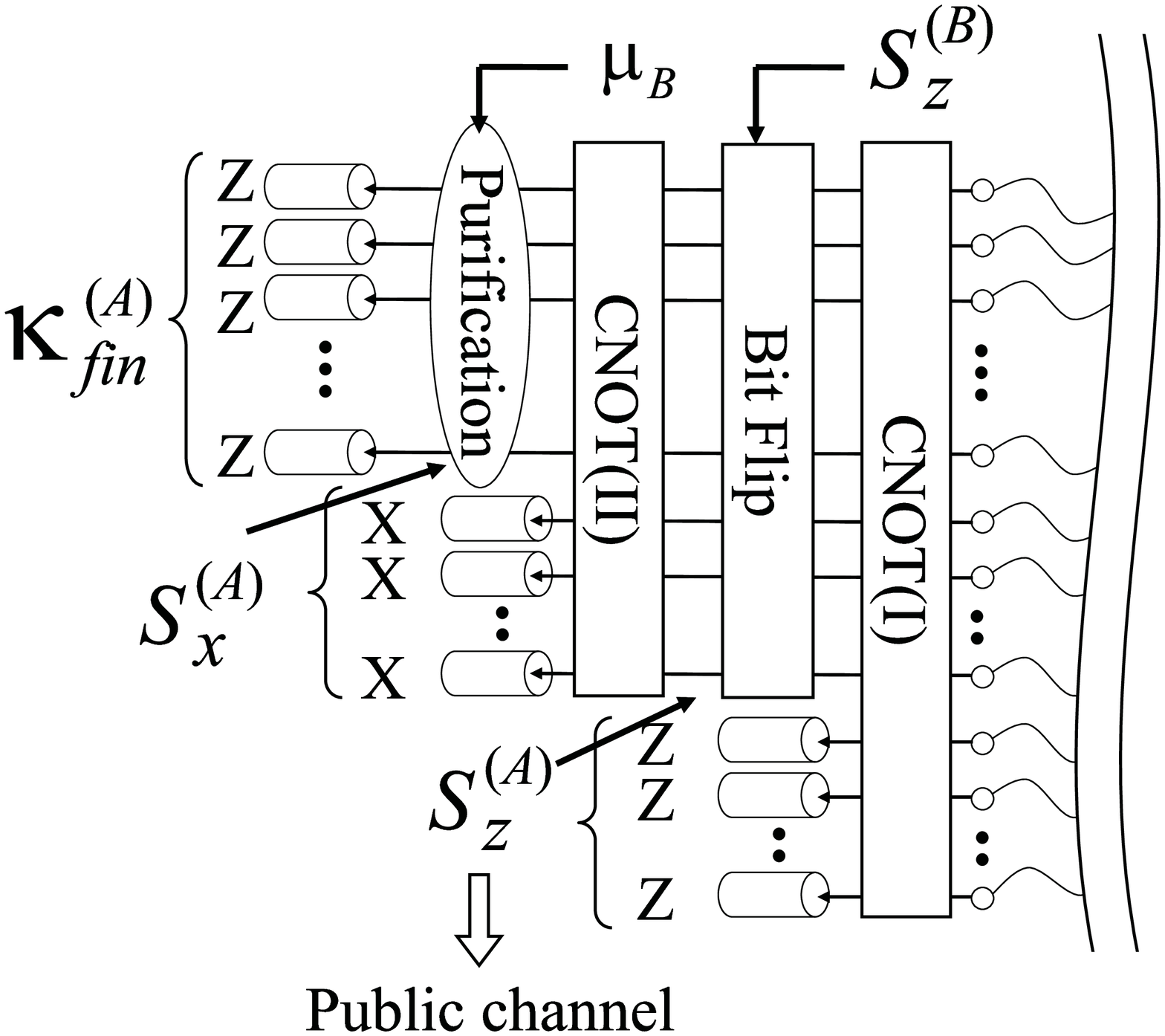}
\end{center}
 \caption{Schematics of our quantum circuit for the {\it Virtual protocol}. ``bit-flip'' consists of the identity and the bit-flip operation $\hat\sigma_x$, and ``Purification'' tries to purify the $(n-m)$ qubits by using $\mu_{B}$ and $S^{(A)}_{x}$.
\label{VP}}
\end{figure}

Note that Step 3 corresponds to the privacy amplification \cite{nielsen}. In order to prove the security of $\kappa_{\rm fin}^{(A)}$ we propose the following virtual protocol (see also Fig. \ref{VP}).

{\it Virtual protocol:}\newline
(Step 1v) Alice prepares the first set of CNOT quantum circuits (we call it CNOT(I)) and applies it to her initial $(n+s)$ qubits. \newline
(Step 2v) Alice conducts $Z$-basis measurements on $s$ qubits of the output ports of CNOT(I) to obtain the syndrome $S^{(A)}_{z}$ \cite{note} and sends it to Bob over a public channel without the encryption. She discards the $s$ measured qubits, and depending on the syndrome from Bob, Alice applies bit-flip operations on some qubits among the remaining $n$ qubits. \newline
(Step 3v) Alice prepares the second set of CNOT quantum circuits (CNOT(II)), and she applies it to her $n$ qubits. Next, Alice generates $m$-bit syndrome in $X$-basis $S^{(A)}_{x}$ by measuring the $m$ qubits of the output port of CNOT(II) along $X$-basis, and after receiving an information $\mu_{B}$ from Bob, Alice obtains $(n-m)$ almost pure and direct product qubit states in $X$-basis. The final key $\kappa_{\rm fin}^{(A)}$ is obtained by performing $Z$-basis measurement on the $(n-m)$ qubits.

Note that CNOT(II) is chosen randomly among the set of CNOT circuits with the following property: the measurement on particular $m$ output ports in $X$-basis is the same as measuring $\{X^{{\vec U}_i}\}_{i=1,2,..m}$ (${\vec U}_i\in{\bf F}_{2}^{n}$) of the $n$ input ports and the measurement on the remaining $n-m$ output ports in $Z$-basis corresponds to the privacy amplification in (Step 3) of the {\it Actual protocol}, i.e., the measurement of $\{Z^{{\vec V}_k}\}_{k=1,2,..n-m}$ (${\vec V}_k\in{\bf F}_{2}^{n}$) on the $n$ input ports. Here, $Z^{\vec x}\equiv Z^{x_1}\otimes Z^{x_2}\otimes\cdots\otimes Z^{x_{n}}$ with $Z^0=\openone$ and $Z^1=Z$ and similar for $X$. The first property of CNOT(II) corresponds to the random hashing along $X$-basis \cite{EDP}, and we note that if we make the hashing random, then the privacy amplification becomes automatically random and vice versa since ${\vec U}_i$ and ${\vec V}_i$ are always orthogonal \cite{note3}.

In (Step 3v) $\mu_{B}$ represents Bob's measurement outcome on his system, which is not necessarily the system of qubits. We assume that $\mu_{B}$ gives Bob the estimation of the $n$ bits of Alice's fictitious $X$-basis measurement outcome ${\bf X}_{A}$ with some uncertainty. The uncertainty will be removed by Alice's random hashing along $X$-basis so that she can distill a direct product of $X$-basis eigenstates and generate the secret key. Here, note that $\mu_{B}$ is not used for the later active quantum operation such as a phase-flip operation. One of the important points in the {\it Virtual protocol} is that since $S^{(A)}_{z}$ and $S^{(A)}_{x}$ are generated via measurements on the different systems, these measurements commute. Similarly, the final measurement along $Z$-basis and the measurement of $S^{(A)}_{x}$ also commute, and the measurement of the final key in the virtual protocol is the same as the one of the actual protocol thanks to the property of CNOT(II). Hence, the newly introduced measurement of $S^{(A)}_{x}$ does not disturb any measurement outcomes in the {\it Actual protocol}, and the {\it Virtual protocol} is not equivalent to the {\it Actual protocol} from Eve's view with respect to $\mu_{B}$ and the measurement of $S_{x}^{(A)}$, which can be removed without degrading the security as we will see.

For the security proof of the {\it Virtual protocol}, we again note that since CNOT(II) is chosen randomly and independently, the measurement of $S^{(A)}_{x}$ serves as a random hashing \cite{EDP} along $X$-basis, hinting that if the number of the rounds of the random hashing is properly chosen, Alice can generate an almost pure and direct product state in $X$-basis. Actually, this is the idea used in the security proof in \cite{Koashi05}, and its sketch of the proof is summarized as follows. The proof starts with putting the assumption that the uncertainty of ${\bf X}_{A}$ from Bob's view after obtaining $\mu_{B}$ is, say $n\xi$ bits. More precisely, we make the following assumption \cite{Koashi05}.

{\it Assumption:} There exists a set $T_{\mu_{B}}^{(n)}$ of $n$-bit sequences with cardinality $|T_{\mu_{B}}^{(n)}|\le2^{n\xi}$ for each $\mu_{B}$, such that the pair of measurement outcomes $(\mu_{B}, {\bf X}_{A})$ satisfies ${\bf X}_{A}\in T_{\mu_{B}}$ except for an exponentially small probability $\eta$. 

By invoking the random hashing argument in \cite{EDP} and by setting $m$ as slightly larger than $n\xi$, we can show that $\bra{t_x}\hat\sigma\ket{t_x}\ge 1-\eta'$ where $\eta'\equiv\eta+2^{-n\epsilon}$. Here, $\ket{t_x}$ is the $n$-qubit $X$-basis state that Alice thinks to have successfully distilled and $\hat\sigma$ is the actually distilled state. Note that the exponentially high fidelity guarantees that $\kappa_{\rm fin}^{(A)}$ is composably secure \cite{composability}.

Next, we convert the {\it Virtual protocol} to the {\it Actual protocol}, i.e., we remove $\mu_{B}$ and $S_{x}^{(A)}$ from the {\it Virtual protocol} while we keep the final key the same. The important point is that we do not use the measurement outcome of $S^{(A)}_{x}$ and $\mu_B$ for any of the following active quantum manipulations, such as bit-flip operations or other quantum evolutions. Moreover, the measurement of $S^{(A)}_{x}$ commutes with the measurement of $\kappa_{\rm fin}^{(A)}$. Thus, even if we skip the measurement of $S_{x}^{(A)}$ and sending $\mu_{B}$, the final key still remains exactly the same, which ends the conversion, and only the fact that Alice could have generated the almost pure state is enough to prove the security.

\section{Applications of our circuit}

So far, we have seen that the security can be proven by making the {\it Assumption}. In what follows, we consider three particular cases to see how to confirm {\it Assumption}. (i) The first case, discussed in Sec. \ref{first case}, is that only Alice has the fictitious qubits and the syndrome is sent from Alice to Bob or bi-directionally \cite{CASCADE}. (ii) The second case, in Sec. \ref{second case}, is that both of Alice and Bob have the qubits, and a bi-directional code is used. (iii) Finally, in Sec. \ref{noisy processing}, we explain, by using our circuit with a shield system, why noisy processing increases the bit error rate threshold. This noisy processing was first introduced by Renner, {\it et. al} \cite{Renner05} and later explained from the viewpoint of $\Gamma$-state distillation \cite{Renes07}.

For the discussion, we assume that thanks to the test bits, Alice and Bob know the rate $e_{p}$ of the ``phase errors'', i.e., Alice's bit value ($\tilde {\bf X}_{A}$), which could have been obtained via the fictitious $X$-basis measurement on the $(n+s)$ qubits (code bits) and Bob's prediction ($\tilde\mu_{B}$) on $\tilde {\bf X}_{A}$ differ. More precisely \cite{nielsen}, for $\epsilon>0$
\begin{eqnarray} 
{\rm P}(|e_{p}-e_{p}^{(t)}|\ge\epsilon)\le O(n+s)2^{-(n+s)\epsilon^2}
\label{test bits}
\end{eqnarray}
holds, where $e_{p}^{(t)}$ is the phase error rate of the test bits. The important point for all the cases (i)-(iii) is that Alice can in principle write down the candidate for ${\bf X}_{A}$ as $T^{(n)}_{{\tilde \mu}_{B}, e_{p}^{(t)}}$ or the candidate for $\tilde {\bf X}_{A}$ as $T^{(n+s)}_{\tilde\mu_{B}, e_{p}^{(t)}}$, given ${\tilde \mu}_{B}$ and $e_{p}^{(t)}$. Another point is that CNOT quantum gates, if we are only interested in a particular basis, can be expressed just as a linear and one-to-one function on the binary bit space. For the explanation, let $F_{CNOT(I)}$ as the linear and one-to-one function on binary $(n+s)$-bit space, which corresponds to CNOT(I) in $X$-basis. 

We remark that when one of the parties, say Bob, has no fictitious qubit, but Alice has, and they use a bi-directional code, we assume that Bob can simultaneously output ${\tilde \mu}_{B}$ and the syndrome for the bit errors $S_{z}^{(B)}$.

\subsection{Only Alice has the fictitious qubit}\label{first case}

Thanks to the estimation and Bob's measurement, Alice can write down the candidate for $\tilde {\bf X}_{A}$ as $T^{(n+s)}_{\tilde\mu_{B}, e_{p}^{(t)}}=\{\tilde\mu_{B}+r_{i}\}_{i=1,2,...,2^{(n+s)h(e_{p}^{(t)}+\epsilon)}}$. Here, $h(x)\equiv -x\log x-(1-x)\log (1-x)$ and $\{r_{i}\}_{i=1,2,...,2^{(n+s)h(e_{p}^{(t)})}}$ represents a set of independent $(n+s)$-bit strings that contain at most $(n+s)(e_{p}^{(t)}+\epsilon)$ 1s. From $T^{(n+s)}_{\tilde\mu_{B}, e_{p}^{(t)}}$, we can calculate the candidate for ${\bf X}_{A}$ as $T^{(n)}_{\tilde\mu_{B}, e_{p}^{(t)}}=\{{\rm Tr}_{\overline n}[F_{CNOT(I)}(\tilde\mu_{B}+r_{i})]\}_{i=1,2,...,2^{(n+s)h(e_{p}^{(t)}+\epsilon)}}$, where ${\rm Tr}_{\overline n}$ means that we discard the last $s$-bit of each $(n+s)$-bit string. Since $F_{CNOT(I)}$ is an one-to-one function, $|T^{(n)}_{\tilde\mu_{B}, e_{p}^{(t)}}|\le2^{(n+s)h(e_{p}^{(t)}+\epsilon)}$ holds, following that $(n+s)(h(e_{p}^{(t)}+\epsilon)+\epsilon')$ rounds of the random hashing is enough for Alice to distill the $X$-basis eigenstate. Here, $\epsilon'$ is a small positive number. We note that the key generation rate $G$ in the limit of large $n$ can be written as
\begin{eqnarray}
G=(n+s)[1-h(e_{p}^{(t)})]-(s+d)\,,
\end{eqnarray} 
where, $d$ is the number of bits that Alice discards when Alice and Bob use bi-directional codes \cite{CASCADE}. Note that the security of Bob's key follows from the direct application of the complementarity scenario \cite{Koashi07}.

\subsection{Both parties have the fictitious qubits}\label{second case}

We assume that Bob has the same quantum circuit as Alice, and Bob measures the first $n$ qubits of the output port of CNOT(I) in $X$-basis, whose outcome is $n$-bit string $\mu_{B}$. Like the case in (i), we can write down the candidate for $\tilde {\bf X}_{A}$ as $T^{(n+s)}_{\mu_{B}, e_{p}^{(t)}}=\{F_{CNOT(I)}^{-1}(\mu_{B}'+a_{k})+r_{i}\}_{i=1,2,...,2^{(n+s)h(e_{p}^{(t)}+\epsilon)},k=1,2,...,2^{s}}$, where $a_{k}$ is arbitrary $2^{s}$ $(n+s)$-bit strings with the first $n$ bits all being zero, and $\mu_{B}'$ is $(n+s)$-bit string with the first $n$ bits being the same as $\mu_{B}$ and all the last $s$ bits being zero. Next, in order to calculate the required quantity $T^{(n)}_{\mu_{B}, e_{p}^{(t)}}$, we apply Alice's $F_{CNOT(I)}$ to each member of $T^{(n+s)}_{\mu_{B}, e_{p}^{(t)}}$ to obtain $T^{(n)}_{\mu_{B}, e_{p}^{(t)}}=\{\mu_{B}+{\rm Tr}_{\overline n}[F_{CNOT(I)}(r_{i})]\}_{i=1,2,...,2^{(n+s)h(e_{p}^{(t)}+\epsilon)}}$ \cite{note2}, where we have used $F_{CNOT(I)}[F_{CNOT(I)}^{-1}(\mu_{B}'+a_{k})]=\mu_{B}$ and the linearity of $F_{CNOT(I)}$. Since $F_{CNOT(I)}$ is an one-to-one function, $|T^{(n)}_{\mu_{B}, e_{p}^{(t)}}|\le2^{(n+s)h(e_{p}^{(t)}+\epsilon)}$ holds, again following that $(n+s)(h(e_{p}^{(t)}+\epsilon)+\epsilon')$ rounds of the random hashing is enough for Alice to distill the $X$-basis eigenstate. We note that the amount of the privacy amplification is determined by $e_{p}^{(t)}$, which means that the security level of Alice's and Bob's keys are the same. Note that the key generation rate $G$ in the limit of large $n$ can be written as
\begin{eqnarray}
G=(n+s)[1-h(e_{p}^{(t)})]-(s+d)\,.
\end{eqnarray} 

\subsection{Noisy processing}\label{noisy processing}

\begin{figure}[tbp]
\begin{center}
 \includegraphics[scale=0.25]{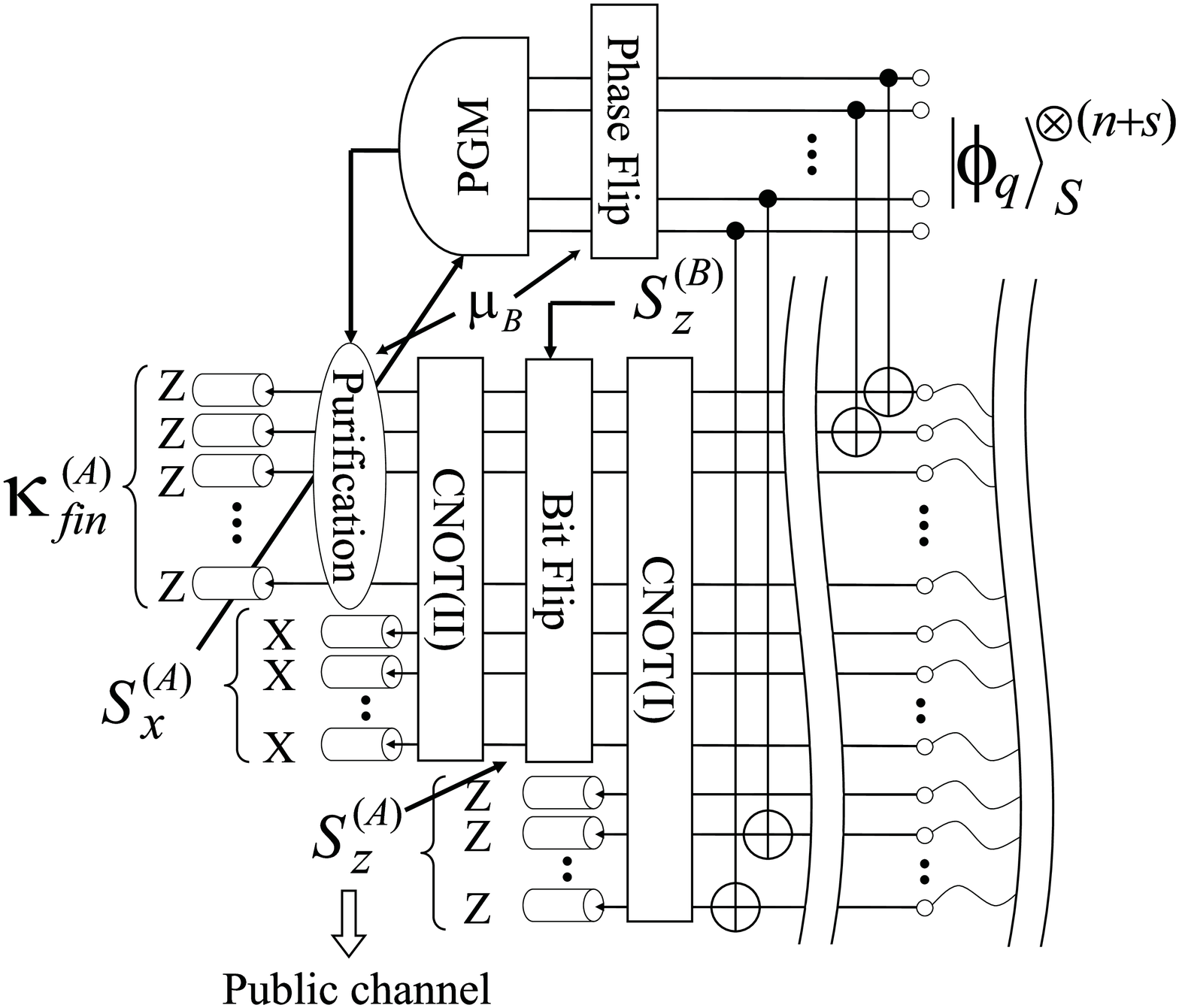}
\end{center}
 \caption{Schematics of the circuit for noisy processing. A difference from the circuit in Fig. \ref{VP} is the shield part whose initial state is $\ket{\phi_q}_{S}^{\otimes (n+s)}$. Each code qubit is connected with each shield via the CNOT gate, and PGM represents the pretty good measurement.
\label{NP}}
\end{figure}

In noisy processing \cite{Renner05}, Alice randomly adds bit errors to her sifted key with probability $0\le q\le1\wedge q\neq 1/2$. This process can be alternatively realized first by preparing shields in the state $\ket{\phi_q}_{S}^{\otimes (n+s)}$ where $\ket{\phi_q}_{S}\equiv\sqrt{1-q}\ket{0_z}_{S}+\sqrt{q}\ket{1_z}_{S}$, and then interacting each of them with Alice's code qubits via CNOT gate with the shields being the control qubit with respect to $Z$-basis (see also Fig. \ref{NP}) \cite{Renes07}. The point is that the role of the target qubit and the control qubit in CNOT is exchanged according to what basis we are working on. Suppose that the state of Alice's code bit before the interaction is a classical mixture of a pure state, say
$\ket{\Psi}_{C}=\sum_{{\vec x}\in T^{(n+s)}_{{\tilde \mu}_{B}, e_p}}\alpha_{{\vec x}}\ket{e_{{\vec x}}}_{C}$ ($\sum_{{\vec x}\in T^{(n+s)}_{{\tilde \mu}_{B}, e_p}}|\alpha_{\vec x}|^2=1$) where ${\vec x}=(x_1, x_2, \cdots, x_{(n+s)})$ is a $(n+s)$-bit string and $\ket{e_{\vec x}}$ is the $X$-basis eigenstate. Then, the CNOT transforms $\ket{\phi_q}_{S}^{\otimes (n+s)}\ket{\Psi}_{C}$ into $\sum_{{\vec x}\in T^{(n+s)}_{{\tilde \mu}_{B}, e_p}}\alpha_{{\vec x}}(Z^{\vec x}\ket{\phi_q}_{S}^{\otimes (n+s)})\ket{e_{{\vec x}}}_{C}$, meaning that the information of Alice's code qubit in $X$-basis is transferredthe  to the shield, and the information is encoded in two nonorthogonal states $\ket{\phi_q}_S$ and $Z\ket{\phi_q}_S$. Thus, by using the information that can be extracted from measuring the shield, Alice can reduce the amount of hashing along $X$-basis, which enhances the bit error rate threshold as well as increases the key generation rate. 

In what follows, we assume that Alice first applies $m$ rounds of the random hashing to narrow down the set of candidates $T_{\tilde \mu_B, e_p}^{(n+s)}$ to a smaller set $\Omega_{m}$, and then she performs the fictitious measurement on the shield to identify the phase error pattern $\tilde {\bf X}_A$. Moreover, in order for $Z\ket{\phi_q}_{S}$ to represent the phase error between Alice and Bob, we apply a phase-flip operation $Z^{{\tilde \mu}_{B}}$ to the shield before measuring it. For simplicity of this discussion, we redefine ${\tilde \mu}_{B}+{\vec x}$ as ${\vec x}$, ${\tilde \mu}_{B}+\tilde{\bf X}_{A}$ as $\tilde{\bf X}_{A}$, and $\{{\tilde \mu}_{B}+{\vec x}\}_{{\vec x}\in\Omega_{m}}$ as $\Omega_{m}$. Furthermore, we define $\ket{\vec x}\equiv Z^{{\vec x}}\ket{\phi_q}_{S}^{\otimes (n+s)}$.

What we have to construct is a positive operator valued measure (POVM) \cite{nielsen} which can identify all the members in $\{Z^{\vec x}\ket{\phi_q}_{S}^{\otimes (n+s)}\}_{{\vec x}\in\Omega_{m}}$ with exponentially small failure probability in $(n+s)$. For the construction, we employ the idea of the so-called Pretty Good Measurement (PGM) \cite{PGM, nielsen}. Originally, PGM can discriminate all the states in a subset of the set stemming from an Identically and Independently Distributed (IID) source, and we cannot directly apply this idea to our case where all the members in the {\it whole set} have to be identified and $\{\vec x\}$ is not stemming from an IID source \cite{comment}. However, observe that $T_{\tilde \mu_B, e_p}^{(n+s)}$ is obtained via the classical random sampling theorem or other estimation method, and this set is very similar to the one stemming from an IID source with the phase error rate $e_p$, i.e., in either case the most likely bit strings are those containing $(n+s)e_p$ 1's. Thus, we infer that if $T_{\tilde \mu_B, e_p}^{(n+s)}$ is contained in the typical space of the IID source, then we can remove the limitation of the IID. Actually, as we will see later, it can be shown that this intuition with a generalized analysis of PGM solves our problem.  

First, we define $\hat \rho$ as $(1-e_{p}^{(t)})\ket{\phi_q}\bra{\phi_q}+e_{p}^{(t)}Z\ket{\phi_q}\bra{\phi_q}Z$, where $e_{p}^{(t)}$ is the phase error rate of the test bit defined in Eq.~(\ref{test bits}) and assume that $\hat \rho$ is diagonalized as
\begin{eqnarray} 
\hat \rho=\lambda_{0}\ket{\tilde 0}\bra{\tilde 0}+\lambda_{1}\ket{\tilde 1}\bra{\tilde 1}.
\end{eqnarray}
We also define $\tilde {\bf a}$ as an $(n+s)$-bit string with respect to the $\{\ket{\tilde 0}, \ket{\tilde 1}\}$ basis, and thanks to the Bernoulli trial argument, we have for $\omega>0$ and $(n+s)>0$ \cite{coheB92},
\begin{eqnarray} 
{\rm P}(||\tilde {\bf a}|-(n+s)\lambda_1|\ge (n+s)\omega)\le2^{1-(n+s)\omega^2}\,,\nonumber\\
\label{omega-typical}
\end{eqnarray}
where $|\tilde {\bf a}|$ represents the Hamming distance of $\tilde {\bf a}$, and we call the space spanned by a set of the $\{\ket{\tilde0}, \ket{\tilde1}\}$-basis eigenstates with $\tilde {\bf a}$ satisfying $||\tilde {\bf a}|-(n+s)\lambda_1|\le (n+s)\omega$ as $\omega$-typical subspace. For later use, we define the projector onto this subspace as $\hat P_{\lambda}^{\omega}$, and note that the random sampling theorem (Eq.~(\ref{test bits})) states that
\begin{eqnarray} 
{\rm P}(||{\vec x}|-(n+s)e_{p}^{(t)}|\ge (n+s)\epsilon)\le O(n+s)2^{-(n+s)\epsilon^2}\,.
\label{epsilon-typical}
\end{eqnarray}
Note that if we chose $\epsilon$ and $\omega$ such that $\epsilon\le\omega$ then the actual phase error pattern $\tilde{\bf X}_{A}$ is included in $\omega$-typical subspace with exponentially close probability to $1$. As we will explain the details in the Appendix, by averaging over the random hashing, POVM $\{\hat M_{\vec x}\}$
\begin{eqnarray}
&&\hat M_{\vec x}\nonumber\\
&=&\left[\sum_{\vec x\in\Omega_{m}}\hat{P}(\hat P_{\lambda}^{\omega}\ket{\vec x})\right]^{-1/2}{\hat P}(\hat P_{\lambda}^{\omega}\ket{\vec x})\left[\sum_{\vec x\in\Omega_{m}}\hat{P}(\hat P_{\lambda}^{\omega}\ket{\vec x})\right]^{-1/2}\,,\nonumber\\
\end{eqnarray}
where $\hat P(\ket{\psi})\equiv\ket{\psi}\bra{\psi}$, we can identify any $\vec x\in\Omega_{m}$ with probability exponentially close to $1$ in $(n+s)$, i.e.,
\begin{eqnarray}
& &E_{\Omega_m}\left(\bra{\vec x}\hat M_{\vec x}\ket{\vec x}\right)\nonumber\\
&\ge&1-6(n+s)2^{-(n+s)\left(\omega^2+\epsilon\log\left(\frac{e_{p}^{(t)}-\epsilon}{1-e_{p}^{(t)}+\epsilon}\cdot\frac{e_{p}^{(t)}}{1-e_{p}^{(t)}}\right)\right)}\nonumber\\
&+&2^{-(n+s)\left[-h(e_{p}^{(t)})+S(\hat\rho)+m/(n+s)-\epsilon-\omega\right]}\,,\nonumber\\
\label{avefail}
\end{eqnarray}
where $E_{\Omega_m}$ represents the averaging over the random hashing, and $S(\hat\rho)\equiv -{\rm Tr}(\hat\rho\log_2\hat\rho)$. By combining Eq.~(\ref{avefail}) with the failure probability of the random sampling for the phase error rate from Eq.~(\ref{epsilon-typical}), the overall failure probability $p_{\rm fail}$ of the identification of the correct phase error pattern can be upper-bounded by
\begin{eqnarray}
p_{\rm fail}&\le&1-E_{\Omega_m}\left(\bra{\vec x}\hat M_{\vec x}\ket{\vec x}\right)+O(n+s)2^{-(n+s)\epsilon^2}\nonumber\\
&\le&6(n+s)\cdot2^{-(n+s)\left(\omega^2+\epsilon\log\left(\frac{e_{p}^{(t)}-\epsilon}{1-e_{p}^{(t)}+\epsilon}\cdot\frac{e_{p}^{(t)}}{1-e_{p}^{(t)}}\right)\right)}\nonumber\\
&+&2^{-(n+s)\left[-h(e_{p}^{(t)})+S(\hat\rho)+m/(n+s)-\epsilon-\omega\right]}\nonumber\\
&+&O(n+s)2^{-(n+s)\epsilon^2}\,.
\end{eqnarray}
Thus, in order for the failure probability to be exponentially small in $(n+s)$, we have to choose parameters such that
\begin{eqnarray}
\omega=\sqrt{\epsilon\log\left(\frac{1-e_{p}^{(t)}+\epsilon}{e_{p}^{(t)}-\epsilon}\cdot\frac{1-e_{p}^{(t)}}{e_{p}^{(t)}}\right)+\delta}\nonumber\\
\label{tomega}
\end{eqnarray}
and 
\begin{eqnarray}
m=(n+s)\left(h(e_{p}^{(t)})-S(\hat\rho)+\epsilon+\omega+\delta_m\right)\,,
\end{eqnarray}
where $\delta$ and $\delta_m$ are small positive numbers. Eq.~(\ref{tomega}) means that $\omega$ has to be larger than $\epsilon$ (assuming $0< \omega\le1$), which confirms our inference. With the above choice of the parameter sets, the fidelity $F$ of the final state $\hat\sigma$ with respect to the target $X$-basis eigenstate $\ket{t_x}$, i.e., $\bra{t_x}\hat\sigma\ket{t_x}$ is expressed as
\begin{eqnarray}
F&=&1-p_{\rm fail}\nonumber\\
&\ge&1-(n+s)6\cdot2^{-(n+s)\delta}-2^{-(n+s)\delta_m}\nonumber\\
&-&O(n+s)2^{-(n+s)\omega^2}\,,
\end{eqnarray}
and the key generation rate $G$ in the limit of large $n$ is
\begin{eqnarray}
G=(n+s)[1-(h(e_{p}^{(t)})-S(\hat\rho))]-(s+d)\,.
\end{eqnarray}
As an example, we consider BB84 and we assume that the phase error rate and unprocessed bit error rates are the same and we use an ideal error correcting code. It follows that $G$ is asymptotically given by 
\begin{eqnarray}
G&\propto&1-[h(e_{p}^{(t)})-S(\hat\rho)]\nonumber\\
&-&h(e_{p}^{(t)}(1-q)+q(1-e_{p}^{(t)}))
\end{eqnarray}
with $e_{p}^{(t)}$ being the same as the bit error rate in the test bits. In this case, it can be seen that the bit error rate threshold is $12.4\%$ with $q\rightarrow1/2$. This value matches the rate provided by \cite{Renner05, Renes07}. Intuitively, when $q$ is close to $1/2$, Alice has to discriminate almost orthogonal states, and a negligibly small amount of privacy amplification is needed. 

Moreover, our analysis can be applied to six-state protocol \cite{Six}. In order to compare the bit error rate threshold given in \cite{Renner05, Renes07}, we assume that Alice and Bob possess the fictitious qubits. In this case, we can employ the mutual information between the bit and phase errors to reduce the amount of hashing along $X$-basis \cite{Lo01, note}. Thus, the phase error rate on the code bit is dependent on whether there is a bit error or not, and we can apply our idea to the cases with and without the bit error separately. The resulting key generation rate $G$ in the limit of large $n$ and $s$ is
\begin{eqnarray}
G&=&(n+s)\left\{1-\left[H(Z|X)-\sum_{i=0}^{1}p(X=i)S(\hat\rho_i)\right]\right\}\nonumber\\
&-&[H(X)+d]\,,
\end{eqnarray}
where $H(X)$($=s$) represents the Shannon entropy for the bit error pattern, $H(Z|X)$ represents the conditional Shannon entropy of the phase error pattern given the bit error pattern, and $p(X)$ is the probability that there is the bit error ($X=0$) or not ($X=1$). Moreover, $\hat\rho_i$ is the density matrix conditional on $X=i$, and it is obtained by replacing $e_{p}^{(t)}$ in $\hat\rho$ with the phase error rate probability conditional on the realization of $X$. By performing the optimization again in terms of the bit error rate threshold by varying $q$, we obtain the improved bit error rate threshold $14.1\%$ when $q\rightarrow1/2$. We note that this rate matches the one given by \cite{Renner05, Renes07}.

\section{Summary}

In summary, we have constructed a quantum circuit for the virtual protocol in the complementarity control to remove the encryption of the syndrome. We applied our circuit to the cases (i) Alice has a fictitious qubit, (ii) both Alice and Bob have fictitious qubits, and (iii) noisy processing. In noisy processing, our proof covers the case where only one party has a qubit, which is a generalization of the original proposal. 

In the analysis of noisy processing, we have to discriminate any state in a set with exponentially small failure probability, and a bit string is encoded in the state and the states are nonorthogonal each other. We have formally shown that PGM can solve this problem, which is a generalization of the original PGM idea in which some states in the set have to be discarded and the states are stemmed from an IID source.

\section{Acknowledgement}

We thank T. Tsurumaru, M. Koashi, and H.-K. Lo for very enlightening discussions. This work was in part supported by National Institute of Information and Communications Technology (NICT) in Japan.

\appendix
\section{}
In this Appendix, we show that $E_{\Omega_m}\left(\bra{\vec x}\hat M_{\vec x}\ket{\vec x}\right)$ is exponentially close to $1$ in $(n+s)$ for any $\vec x\in\Omega_m$. In the analysis, we assumed that $0\le e_{p}^{(t)}\le 1/2$, and we used some techniques from \cite{nielsen}. First, one can show that
\begin{eqnarray}
\sum_{{\vec x}\in\Omega_m}\hat M_{\vec x}=\openone_{\{\hat P_{\lambda}^{\omega}\ket{\vec x}\}_{{\vec x}\in\Omega_m}}\le\hat P_{\lambda}^{\omega}\,,
\end{eqnarray}
where $\openone_{\{\hat P_{\lambda}^{\omega}\ket{\vec x}\}_{{\vec x}\in\Omega_m}}$ is the identity operator of the space spanned by $\{\hat P_{\lambda}^{\omega}\ket{\vec x}\}_{{\vec x}\in\Omega_m}$. This means that $\{\hat M_{\vec x}\}$ and an additional positive operator that corresponds to the failure discrimination of a state on $\omega$-typical subspace form a POVM on $\omega$-typical subspace.

Next, a direct calculation shows that
\begin{eqnarray}
& &E_{\Omega_m}\left(\bra{\vec x}\hat M_{\vec x}\ket{\vec x}\right)\nonumber\\
&=&E_{\Omega_m}\left(\bra{\vec x}(\hat P_{\lambda}^{\omega}\hat M_{\vec x}\hat P_{\lambda}^{\omega})\ket{\vec x}\right)\nonumber\\
&=&E_{\Omega_m}\left(\left[\bra{\vec x}\hat P_{\lambda}^{\omega}\left(\sum_{{\vec x}'\in\Omega_{m}}P_{\lambda}^{\omega}\ket{{\vec x}'}\bra{{\vec x}'}P_{\lambda}^{\omega}\right)^{-1/2}\hat P_{\lambda}^{\omega}\ket{\vec x}\right]^2\right)\nonumber\\\label{a1}
\end{eqnarray}
Let us regard $(\bra{{\vec x}}\hat P_{\lambda}^{\omega})\left(\sum_{{\vec x}\in\Omega_{m}}P_{\lambda}^{\omega}\ket{\vec x}\bra{\vec x}P_{\lambda}^{\omega}\right)^{-1/2}(\hat P_{\lambda}^{\omega}\ket{{\vec x}'})$ as the $(x, x')$ component of the matrix $\Gamma^{1/2}$, which is a real value, and by using $y^2\ge2y-1$ for real $y$ and $\Gamma^{1/2}\ge3\Gamma/2-\Gamma^2/2$ \cite{nielsen}, we have
\begin{eqnarray}
(\Gamma^{1/2}|_{x,x})^2&\ge&2\Gamma^{1/2}|_{x,x}-1\nonumber\\
&\ge&3\Gamma|_{x,x}-\Gamma^2|_{x,x}-1\,.
\end{eqnarray}
Moreover, by using
\begin{eqnarray}
\Gamma |_{x,x'}=\sum_{y}(\Gamma^{1/2})_{x,y}(\Gamma^{1/2})_{y,x'}=\bra{\vec x}P_{\lambda}^{\omega}\ket{{\vec x}'},
\end{eqnarray}
we have
\begin{eqnarray}
& &E_{\Omega_m}\left(\bra{\vec x}\hat M_{\vec x}\ket{\vec x}\right)\nonumber\\
&\ge&3\bra{\vec x}P_{\lambda}^{\omega}\ket{\vec x}-E_{\Omega_m}\left(\bra{\vec x}P_{\lambda}^{\omega} \sum_{{\vec x}'\in\Omega_{m}}\ket{{\vec x}'}\bra{{\vec x}'}P_{\lambda}^{\omega}\ket{\vec x}\right)-1\,.\nonumber\\
\label{a2}
\end{eqnarray}
In order to evaluate $\bra{\vec x}P_{\lambda}^{\omega}\ket{\vec x}$ we consider $\bra{\vec x}(\openone-P_{\lambda}^{\omega})\ket{\vec x}$, which is upper-bounded as
\begin{eqnarray}
& &\bra{\vec x}(\openone-P_{\lambda}^{\omega})\ket{\vec x}\nonumber\\
&=&(_{(n+s)}C_{|\vec x|})^{-1}\sum_{{\vec \xi}\,\,{\rm s.t}\,\,|\xi|=|\vec x|}\bra{\vec \xi}(\openone-P_{\lambda}^{\omega})\ket{\vec \xi}\nonumber\\
&\le&\frac{\sum_{{\vec v}}(e_{p}^{(t)})^{|\vec v|}(1-e_{p}^{(t)})^{(n+s)-|{\vec v}|}\bra{\vec v}(\openone-P_{\lambda}^{\omega})\ket{\vec v}}{(_{(n+s)}C_{|\vec x|})(e_{p}^{(t)})^{|\vec x|}(1-e_{p}^{(t)})^{(n+s)-|{\vec x}|}}\,,\nonumber\\
\end{eqnarray}
where the last summation is taken with respect to all $(n+s)$-bit string $\vec v$, and $\openone$ is the identity matrix of the space spanned by all $\ket{\vec v}$. Since $\hat\rho^{\otimes (n+s)}=\sum_{{\vec v}}(e_{p}^{(t)})^{|\vec v|}(1-e_{p}^{(t)})^{(n+s)-|{\vec v}|}\ket{\vec v}\bra{\vec v}$ holds, we have
\begin{eqnarray}
& &\bra{\vec x}(\openone-P_{\lambda}^{\omega})\ket{\vec x}\nonumber\\
&\le&\frac{{\rm Tr}\left(\hat\rho^{\otimes (n+s)}(\openone-P_{\lambda}^{\omega})\right)}{(_{(n+s)}C_{|\vec x|})(e_{p}^{(t)})^{|\vec x|}(1-e_{p}^{(t)})^{(n+s)-|{\vec x}|}}.\nonumber\\
\end{eqnarray}
Note that
\begin{eqnarray}
&&(_{(n+s)}C_{|\vec x|})(e_{p}^{(t)})^{|\vec x|}(1-e_{p}^{(t)})^{(n+s)-|{\vec x}|}\nonumber\\
&\ge&\frac{2^{(n+s)h(\frac{|\vec x|}{(n+s)})+|\vec x|\log_2e_{p}^{(t)}+(n+s-|{\vec x}|)\log_2(1-e_{p}^{(t)})}}{(n+s)}\,,\nonumber\\
\end{eqnarray}
and its exponent is lower-bounded as
\begin{eqnarray}
&&(n+s)h\left(\frac{|\vec x|}{(n+s)}\right)+|\vec x|\log_2e_{p}^{(t)}\nonumber\\&+&(n+s-|{\vec x}|)\log_2(1-e_{p}^{(t)})\nonumber\\
&=&(n+s)\left[h\left(\frac{|\vec x|}{(n+s)}\right)-h(e_{p}^{(t)})\right]\nonumber\\
&-&\left[(n+s)e_{p}^{(t)}-|{\vec x}|\right]\log_2\frac{e_{p}^{(t)}}{1-e_{p}^{(t)}}\nonumber\\
&\ge&(n+s)\epsilon\left[-h^{\prime}\left(e_{p}^{(t)}-\epsilon\right)+\log\frac{e_{p}^{(t)}}{1-e_{p}^{(t)}}\right]\nonumber\\
&=&(n+s)\epsilon\log\left(\frac{e_{p}^{(t)}-\epsilon}{1-e_{p}^{(t)}+\epsilon}\cdot\frac{e_{p}^{(t)}}{1-e_{p}^{(t)}}\right)\,.
\end{eqnarray}
On the other hand, from Eq.~(\ref{omega-typical}), we have 
\begin{eqnarray}
{\rm Tr}\left(\hat\rho^{\otimes (n+s)}(\openone-P_{\lambda}^{\omega})\right)\le2^{1-(n+s)\omega^2}\,,
\end{eqnarray}
following that $\bra{\vec x}P_{\lambda}^{\omega}\ket{\vec x}$ is lower-bounded by
\begin{eqnarray}
& &\bra{\vec x}P_{\lambda}^{\omega}\ket{\vec x}\nonumber\\
&\ge&1-(n+s)2^{1-(n+s)\left(\omega^2+\epsilon\log\left(\frac{e_{p}^{(t)}-\epsilon}{1-e_{p}^{(t)}+\epsilon}\cdot\frac{e_{p}^{(t)}}{1-e_{p}^{(t)}}\right)\right)}\,.\nonumber\\
\label{typicalerror}
\end{eqnarray}

Next, we evaluate the second term of Eq.~(\ref{a2})
\begin{eqnarray}
&&E_{\Omega_m}\left(\bra{\vec x}P_{\lambda}^{\omega} \sum_{{\vec x}'\in\Omega_{m}}\ket{{\vec x}'}\bra{{\vec x}'}P_{\lambda}^{\omega}\ket{\vec x}\right)\nonumber\\
&=&E_{\Omega_m}\left(\bra{\vec x}P_{\lambda}^{\omega} \sum_{{\vec y}\in T^{(n+s)}_{\mu_{B}, e_p}}\theta_{\Omega_m}({\vec y})\ket{\vec y}\bra{\vec y}P_{\lambda}^{\omega}\ket{\vec x}\right)\nonumber\\
&=&\bra{\vec x}P_{\lambda}^{\omega} \sum_{{\vec y}\in T^{(n+s)}_{\mu_{B}, e_p}}2^{-m}\ket{\vec y}\bra{\vec y}P_{\lambda}^{\omega}\ket{\vec x}\nonumber\\
&+&(1-2^{-m})\left|\bra{\vec x}P_{\lambda}^{\omega}\ket{\vec y}\right|^2\nonumber\\
&\le&\bra{\vec x}P_{\lambda}^{\omega} \sum_{{\vec y}\in T^{(n+s)}_{\mu_{B}, e_p}}2^{-m}\ket{\vec y}\bra{\vec y}P_{\lambda}^{\omega}\ket{\vec x}+1\,,\nonumber\\
\end{eqnarray}
where $\theta_{\Omega_m}({\vec y})=1$ if ${\vec y}\in\Omega_m$, otherwise $\theta_{\Omega_m}({\vec y})=0$. In the second equality, we used the fact that the bit string corresponding to the actual error pattern ${\tilde X}_{A}\in T_{{\tilde \mu}_{B}, e_p}^{(n+s)}$ is always selected as $\Omega_{m}$, and the probability that ${\vec y}\neq {\tilde X}_{A}\wedge{\vec y}\in T_{{\tilde \mu}_{B}, e_p}^{(n+s)}$ is chosen as $\Omega_m$ is $2^{-m}$. Observe that
\begin{eqnarray}
2^{-(n+s)(h(e_{p}^{(t)})+\epsilon)}\sum_{{\vec y}\in T^{(n+s)}_{\mu_{B}, e_p}}\ket{\vec y}\bra{\vec y}\le\hat\rho^{\otimes (n+s)} \,,
\end{eqnarray}
and 
\begin{eqnarray}
&&\bra{\vec x}P_{\lambda}^{\omega}\hat\rho^{\otimes (n+s)}P_{\lambda}^{\omega}\ket{\vec x}\nonumber\\
&\le& \bra{\omega_{\rm typ}}\hat\rho^{\otimes (n+s)}\ket{\omega_{\rm typ}}\nonumber\\
&\le&2^{-(n+s)\left[S(\hat\rho)-\omega\right]}\,,
\end{eqnarray}
where $\ket{\omega_{\rm typ}}$ is a state on the $\omega$-typical subspace. Using them, we have
\begin{eqnarray}
&&E_{\Omega_m}\left(\bra{\vec x}P_{\lambda}^{\omega} \sum_{{\vec x}'\in\Omega_{m}}\ket{{\vec x}'}\bra{{\vec x}'}P_{\lambda}^{\omega}\ket{\vec x}\right)\nonumber\\
&\le&2^{-(n+s)\left[-h(e_{p}^{(t)})+S(\hat\rho)+m/(n+s)-\epsilon-\omega\right]}+1\,.
\end{eqnarray}
Combining all together, we have the final expression as
\begin{eqnarray}
& &E_{\Omega_m}\left(\bra{\vec x}\hat M_{\vec x}\ket{\vec x}\right)\nonumber\\
&\ge&1-6(n+s)2^{-(n+s)\left(\omega^2+\epsilon\log\left(\frac{e_{p}^{(t)}-\epsilon}{1-e_{p}^{(t)}+\epsilon}\cdot\frac{e_{p}^{(t)}}{1-e_{p}^{(t)}}\right)\right)}\nonumber\\
&+&2^{-(n+s)\left[-h(e_{p}^{(t)})+S(\hat\rho)+m/(n+s)-\epsilon-\omega\right]}\,.\nonumber\\
\end{eqnarray}


\begin{thebibliography}{99}
\bibitem{note6} So far, it is known that we can exploit the mutual information between the bit and phase errors when both parties possess a qubit state, which may be generalized in the future to the case where only one party possesses a qubit.
\bibitem{note} In other words, CNOT(I) is designed in such a way that the measurement on the $s$ qubits is equivalent to obtain $S^{(A)}_{z}$.
\bibitem{comment} In \cite{Renes07}, the authors claim that the direct application of the original PGM idea combined with the argument of type is enough to prove the security against coherent attacks.
\bibitem{note} In the six-state protocol, the diagonal elements of the marginal density matrix $\hat\sigma$ for the code qubit pair possessed by Alice and Bob is represented by $(1-3\xi, \xi, \xi, \xi)$, Here, $0\le\xi\le1/3$, and each entry represents the four possibilities: a) no error, b) bit-flip error, but no phase-flip error, c) phase-flip error, but no bit-flip error, and d) both bit and phase-flip errors. The fact that $\hat\sigma$ has the above form implies that there is a nontrivial mutual information between the phase and bit error pattern, as explained in \cite{Lo01}.
\bibitem{note3} As we will state in the next paragraph, the measurements of $S^{(A)}_{z}$ and $S^{(A)}_{x}$ commute. This means that $\{X^{{\vec U}_i}\}_{i=1,2,..m}$ commutes with $\{Z^{{\vec V}_k}\}_{k=1,2,..n-m}$, following that ${\vec U}_i\cdot{\vec V}_k=0$.
\end{thebibliography}
\end{document}